\documentclass[11pt]{CGC2}

\usepackage[english]{babel}
\usepackage{verbatim}
\usepackage{graphicx}
\usepackage[english]{babel}
\usepackage{amsmath}
\usepackage{amssymb}
\usepackage{amsthm}
\usepackage[mathcal]{eucal}
\usepackage[latin1]{inputenc}

\usepackage{url}
\usepackage{hyperref}
\usepackage{breakurl}

\begin{document}

%%%%%%%%%%%%%%%%%%%%%%%%%%%%%%%%%%%%%%%%%%%%%%%%%%%%%%%%%%%%%%%%%%%%%
\Logo{}

\begin{frontmatter}

\title{A bound on the size of linear codes}

{\author{Eleonora Guerrini}}
{{\tt (eleonora.guerrini@ens-lyon.fr)}}\\
{{Ecole Normale Superieure de Lyon, France.}}

{\author{Massimiliano Sala}} {\tt{(maxsalacodes@gmail.com)}}\\
{Department of Mathematics, University of Trento, Italy.}

\runauthor{E.~Guerrini, M.~Sala}

\section*{New version}

A new version of this preprint \emph{``A bound on the size of linear codes (E. Guerrini, M. Sala)''} is available following this link:

\url{arxiv.org/submit/501864/view}

under the title \emph{``A bound on the size of linear codes and systematic codes (E. Guerrini, E. Bellini, M. Sala)''}.

\end{frontmatter}
%==================================================================

%%%%%%%%%%%%%%%%%%%%%%%%%%%%%%%%%%%%%%%%%%%%%%%%%%%%%%%%%%%%%%%%%%%%%%%%%%%%%%%%%%%%

\end{document}